\documentclass[traditabstract]{aa}  
\usepackage{graphicx}
\usepackage{textcomp}
\usepackage{lscape}
\usepackage[varg]{txfonts}
\usepackage{mathrsfs}
\usepackage{amssymb}
\usepackage{natbib}
\usepackage[colorlinks, citecolor={blue}]{hyperref}
\usepackage{color}
\authorrunning{S.~Duarte~Puertas et al.}
\titlerunning{Aperture-free SFR of SDSS star-forming galaxies}

\begin{document} 

   \title{Aperture-free star formation rate of SDSS star-forming galaxies\thanks{SFR for $\sim$210,000 SDSS star-forming galaxies and related relevant data will be made available in electronic form at the CDS via anonymous ftp to cdsarc.u-strasbg.fr (130.79.128.5) or via \texttt{http://cdsweb.u-strasbg.fr/cgi-bin/qcat?J/A+A/}.}}

\author{
S.~Duarte~Puertas\inst{1}
\and 
J.~M.~Vilchez\inst{1}
\and 
J.~Iglesias-P\'{a}ramo\inst{1,2}
\and
C.~Kehrig\inst{1}
\and
E.~P\'{e}rez-Montero\inst{1}
\and
F.~F.~Rosales-Ortega\inst{3}
}
\institute{
Instituto de Astrof\'{\i}sica de Andaluc\'{\i}a - CSIC, Glorieta de la Astronom\'{\i}a s.n., 18008 Granada, Spain\label{inst1} \\ \email{salvini@iaa.es}
\and
and Estaci\'{o}n Experimental de Zonas \'{A}ridas - CSIC, Ctra. de Sacramento s.n., La Ca\~{n}ada, Almer\'{\i}a, Spain\label{inst2}
\and
Instituto Nacional de Astrof\'{\i}sica, \'{O}ptica y Electr\'{o}nica, Luis E. Erro 1, 72840 Tonantzintla, Puebla, Mexico\label{inst3}
}

   \date{Received \today; accepted \today}

 \abstract 
% 5 {} token are mandatory
{Large area surveys with a high number of galaxies observed have undoubtedly marked a milestone in the understanding of several properties of galaxies, such as star-formation history, morphology, and metallicity. However, in many cases, these surveys provide fluxes from fixed small apertures (e.g. fibre), which cover a scant fraction of the galaxy, compelling us to use aperture corrections to study the global properties of galaxies. In this work, we derive the current total star formation rate (SFR) of Sloan Digital Sky Survey (SDSS) star-forming galaxies, using an empirically based aperture correction of the measured $\rm H\alpha$ flux for the first time, thus minimising the uncertainties associated with reduced apertures. All the $\rm H\alpha$ fluxes have been extinction-corrected using the $\rm H\alpha/H\beta$ ratio free from aperture effects. The total SFR for $\sim$210,000 SDSS star-forming galaxies has been derived applying pure empirical $\rm H\alpha$ and $\rm H\alpha/H\beta$ aperture corrections based on the Calar Alto Legacy Integral Field Area (CALIFA) survey. We find that, on average, the aperture-corrected SFR is $\sim$0.65dex higher than the SDSS fibre-based SFR. The relation between the SFR and stellar mass for SDSS star-forming galaxies (SFR--$\rm M_\star$) has been obtained, together with its dependence on extinction and $\rm H\alpha$ equivalent width. We compare our results with those obtained in previous works and examine the behaviour of the derived SFR in six redshift bins, over the redshift range $\rm 0.005 \leq z\leq 0.22$. The SFR--$\rm M_\star$ sequence derived here is in agreement with selected observational studies based on integral field spectroscopy of individual galaxies as well as with the predictions of recent theoretical models of disc galaxies.}


   \maketitle
%
%________________________________________________________________
%% 1. Introduction %%%%%%%%%%%%%%%%%%%%%%
\section{Introduction}
In the last two decades we have witnessed the revolutionary appearance of large area surveys with a huge number of galaxies observed (e.g. Sloan Digital Sky Survey (SDSS), \citealp{2000AJ....120.1579Y}; 2dFGRS, \citealp{2001MNRAS.328.1039C}; VVDS, \citealp{2005A&A...439..845L}; GAMA, \citealp{2011MNRAS.413..971D}). These surveys have been very useful for performing a complete study of the non-uniform distribution of galaxies in the Universe (large-scale structure) and to understand galaxy formation and evolution. They also give us information about important properties of galaxies, such as morphology, stellar mass, star formation rate, metallicity, and dependence on the environment. These surveys used single-fibre spectroscopy with small apertures (e.g. 2'' diameter for 2dFGRS and GAMA, and 3'' diameter for SDSS) and therefore cover a limited region of the galaxy, thus providing partial information on the extensive properties of galaxies. For a more detailed analysis of the integrated properties of each galaxy, it is necessary to make use of integral field spectrographs \citep[IFS, e.g.][]{2012A&A...540A..11K,2016MNRAS.459.2992K}. IFS surveys, such as SAURON \citep{2001MNRAS.326...23B}, PINGS \citep{2010MNRAS.405..735R}, MASSIV \citep{2012A&A...539A..91C}, CALIFA \citep{2012A&A...538A...8S,2015A&A...576A.135G}, and MANGA \citep{2015ApJ...798....7B}, are based on arrays of fibres and allow us to obtain information from the whole galaxy. However, the integration times necessary to observe each galaxy are long, hindering the acquisition of a large number of galaxies in comparison with that obtained from single-fibre surveys.

It is clear that to gain a better insight into the global properties of galaxies we need tools that allow us to link both single-fibre and IFS surveys. In this work we make use of one of these tools for the purpose of studying the total current star formation rate (hereafter SFR) of star-forming galaxies, using their total $\rm H\alpha$ fluxes emitted by the gas ionised by young massive stars \citep[e.g.][]{1998ApJ...498..541K,2009ApJ...703.1672K}. The SFR presents a well-known characteristic relation with the stellar mass \citep[e.g.][]{2004MNRAS.351.1151B}. In the SFR--stellar mass plane (SFR--$\rm M_\star$), active star-forming galaxies define a distinct sequence called main sequence \citep{2007ApJ...660L..43N}. In addition, less active star-forming galaxies (e.g. quenched and ageing star-forming galaxies) appear in this plane located at higher masses and lower SFR values \citep[e.g.][]{2015ApJ...801L..29R,2015MNRAS.451..888C,2016MNRAS.455L..82L}, as well as a family of outliers to this main sequence (MS) that are generally interpreted as starbursts driven by merging \citep[e.g.][]{2011ApJ...739L..40R}. The MS of star formation has been parametrized by the following equation:
\begin{equation} \label{Eq:eq_1}
\rm log(SFR)=\alpha\,log(M_\star)+\beta.          
\end{equation}

From a theoretical point of view, recent studies have obtained values for $\rm \alpha$ (MS slope) near to unity \citep[e.g.][]{2010MNRAS.405.1690D,2015MNRAS.447.3548S,2016MNRAS.456.2982T}. \cite{2010MNRAS.405.1690D} used a semi-analytic model of disc galaxies parametrizing several properties: for example, SFRs and metallicities were computed in a spatially resolved way as a function of galactic radius. For galaxies with stellar masses between $\rm 10^9\ M_\odot$ and $\rm 10^{11}\ M_\odot$ the slope found in \cite{2010MNRAS.405.1690D} is $\rm \alpha =$ 0.96. In contrast, \cite{2015MNRAS.447.3548S} used Illustris \citep[state-of-the-art cosmological hydrodynamical simulation of galaxy formation;][]{2015A&C....13...12N} to reproduce the observed star formation MS, finding a similar slope to \cite{2010MNRAS.405.1690D} at low redshift.
From observational studies quoting $\rm H\alpha$ based SFR determinations, the MS slope varies preferentially between $\sim$0.6 and $\sim$1, and $\rm \beta$ between $\sim$ -9 and $\sim$ -3, depending on the precise methodology and data used \citep[see e.g.][and references therein]{2011ApJ...739L..40R,2014ApJS..214...15S}. An important factor behind this spread in $\rm \alpha$ and $\rm \beta$ values can be related to the aperture corrections applied to the $\rm H\alpha$ measurements.

In order to evaluate the SFR of a galaxy in single-fibre surveys, an aperture correction needs to be applied to account for the entire galaxy. This correction becomes essential in order to analyse the SFR dependence with redshift (z), especially for samples of star-forming galaxies at low z. Many such studies have dealt with the SFR of star-forming galaxies in the local Universe \citep[e.g.][]{2004MNRAS.351.1151B,2006ApJS..164...38I,2007ApJS..173..267S,2008ApJS..178..247K,2010ApJ...721..193P} and others extended to medium and large redshift \citep[e.g.][]{1996MNRAS.283.1388M,2007A&A...468...33E,2010ApJ...721..193P,2012ApJ...754L..29W,2013MNRAS.433..796D,2015MNRAS.454.2015D}. At lower redshift the effects produced by a fixed aperture size will be clearly more significant than for high redshifts. It is important to note that for galaxies about the size of the Milky Way, the 3-arcsec-diameter SDSS fibre never encompasses the complete galactic disc. For this reason, it is always necessary to use aperture corrections to derive the total SFR when using the SDSS $\rm H\alpha$ fluxes. An extra drawback produced by this limitation of the SDSS fibre is particularly relevant for "late-type" spiral galaxies \citep[i.e. disc-dominated galaxies with Hubble type from Sb to Sdm,][]{1997PASP..109.1298D} since they present higher star formation in the outer parts of their discs. Therefore, the $\rm H\alpha$ flux measured by SDSS for these galaxies would lead to an underestimation of the total SFR.

Several studies have already emphasised the importance of quantifying the effect of aperture in the observational data \citep[e.g.][]{2005PASP..117..227K,2008ApJS..178..247K,2014A&A...561A.129M} and also when comparing with the theoretical model predictions \citep{2016MNRAS.462.2046G}. To our knowledge, a solid empirical aperture correction has not yet been implemented in a systematic way for the analysis of large samples of galaxies. Up to now, most studies quantifying the total SFR of galaxies from single-fibre surveys apply aperture corrections using model-based methods \citep[e.g.][]{2004MNRAS.351.1151B,2007ApJS..173..267S}. Other works apply geometrical considerations in order to compensate for the unobserved $\rm H\alpha$ emission of the galaxy, scaling it according to its broad-band photometric map, or by using analytical recipes \citep[e.g.][]{2003ApJ...599..971H,2013MNRAS.430.2047H}. For SDSS star-forming galaxies, \cite{2004MNRAS.351.1151B} originally corrected SDSS fibre SFRs from aperture effects using the resolved colour information available for each galaxy. In the Max-Planck-Institut für Astrophysik and Johns Hopkins University (MPA-JHU) database\footnote{Available at \href{http://www.mpa-garching.mpg.de/SDSS/}{\texttt{http://www.mpa-garching.mpg.de/SDSS/}}.} \citep{2003MNRAS.341...33K,2004MNRAS.351.1151B,2004ApJ...613..898T,2007ApJS..173..267S} the \cite{2004MNRAS.351.1151B} methodology, improved following \cite{2007ApJS..173..267S}, was used to derive SFRs. It is important to note that MPA-JHU galactic SFRs always include the nuclear emission. In addition, the median profiles of the growth curve corresponding to the $\rm H\alpha/H\beta$ aperture correction decreases when galaxy radius increases \citep{2013A&A...553L...7I}. Thus, galaxies for which only the central zones are observed present overestimated extinction \citep[][hereafter IP16]{2016ApJ...826...71I}: on average, some bias is expected in MPA-JHU SFRs since galaxies' extinction gradients were not accounted for \citep{2016MNRAS.455.2826R}.

A rigorous methodology to derive the total SFR of a galaxy should make use of its entire $\rm H\alpha$ flux and $\rm H\alpha/H\beta$ ratio. For single-fibre surveys, the total $\rm H\alpha$ flux of a galaxy can be obtained using an empirical aperture correction for $\rm H\alpha$ derived from IFS of nearby galaxies. According to \citet{2013A&A...553L...7I,2016ApJ...826...71I}, the CALIFA project allows an accurate aperture correction for $\rm H\alpha$ to be derived empirically. IP16 provide the growth curve of $\rm H\alpha$ flux as a function of $\rm R_{50}$, the Petrosian radius containing 50\% of the total galaxy flux in the r-band, on the basis of a representative sample of 165 star-forming galaxies from the CALIFA survey \citep{2012A&A...538A...8S,2013A&A...549A..87H,2014A&A...569A...1W}. In this work we make use of the empirical aperture correction from IP16 in order to obtain the total $\rm H\alpha$ flux for a sample of $\sim$210,000 star-forming galaxies from the SDSS. From this total $\rm H\alpha$ flux we derive aperture-corrected values of SFR in each galaxy, which will allow us to study the global relation between the SFR, stellar mass, and redshift.

To our knowledge this work represents the first attempt to analyse the behaviour of the present-day SFR for a large sample including all SDSS star-forming galaxies using the total $\rm H\alpha$ emission empirically corrected by $\rm H\alpha$ aperture coverage in a systematic manner. The structure of this paper is organised as follows. In Sect.~\ref{sec:data} we describe the data and provide a description of the methodology used to select the star-forming galaxies. We detail the methodology used to derive the entire $\rm H\alpha$ flux using the empirical aperture correction and the corresponding SFR in Sect.~\ref{sec:metho}. Our main results are presented in Sect.~\ref{sec:results} and the associated discussion in Sect.~\ref{sec:discu}. Finally, a summary and the main conclusions of our work are given in Sect.~\ref{sec:conclu}. 

Throughout the paper, we assume a Friedman-Robertson-Walker cosmology with $\Omega_{\Lambda 0}=0.7$, $\Omega_{\rm m 0}=0.3$, and $\rm H_0=70\,km\,s^{-1}\,Mpc^{-1}$. We use the \cite{2001MNRAS.322..231K} universal initial mass function (IMF).\footnote{It is necessary to multiply the \cite{2001MNRAS.322..231K} SFR estimation by 1.6 to transform from the \cite{2001MNRAS.322..231K} IMF to the \cite{1955ApJ...121..161S} IMF and by 0.943 to transform from \cite{2001MNRAS.322..231K} to \cite{2003ApJ...586L.133C} IMF \citep{2009ApJ...701.1765M,2010MNRAS.408.2115M}.}

%__________________________________________________________________
%% 2. Data and sample %%%%%%%%%%%%%%%%%%%%%%
\section{Data and sample}
\label{sec:data}

%__________________________________________________________________
%% 2.1. Samples selection %%%%%%%%%%%%%%%%%%%%%%
\subsection{Sample selection}
\label{sec:samples}
Our study is based on the MPA-JHU public catalogue \citep{2003MNRAS.341...33K,2004MNRAS.351.1151B,2004ApJ...613..898T,2007ApJS..173..267S}, which gives spectroscopic data of the galaxies in the Sloan Digital Sky Survey Data Release 7 (SDSS-DR7) \citep{2009ApJS..182..543A}. The SDSS spectroscopic primary sample of galaxies is complete in Petrosian r magnitude in the range $\rm 14.5\,\leq\,r\,\leq\,17.7$ \citep{2002AJ....124.1810S,2004astro.ph..6220B}. We added the photometric data from the SDSS-DR12 \citep{2000AJ....120.1579Y,2015ApJS..219...12A} to the MPA-JHU catalogue. The MPA-JHU catalogue contains 1,477,411 objects, of which 933,310 are galaxies with spectroscopic properties. In this catalogue, the stellar masses are estimated using spectral energy distribution (SED) fitting to ugriz photometry as described in \cite{2003MNRAS.341...33K}. The line fluxes\footnote{MPA-JHU re-normalised its flux outputs to match the photometric fibre magnitude in the r-band. See the MPA-JHU website.} are corrected for foreground Galactic reddening and extinction using the attenuation curve from \cite{1994ApJ...422..158O} and the extinction from \cite{1998ApJ...500..525S}. Finally, MPA-JHU provides the emission line fluxes measured using Gaussian fittings over the subtracted continuum and the signal-to-noise (S/N) in the whole spectrum. For those galaxies with multiple entries we selected only those with the largest S/N in the whole spectrum, reducing the total number of galaxies to 874,701. 

We selected our primary sample according to the following criteria:

\begin{enumerate}[i)]
\item The galaxy stellar mass [$\rm log(M_\star/M_\odot)$] is selected in the range between 8.50 and 11.50. Galaxies without any estimation of the stellar mass are discarded in order to carry out a proper comparison between the SFR and the stellar mass.

\item We restricted our sample to the galaxies with small relative error of size measurements of $\rm R_{50}$ ($\rm \Delta(R_{50})/R_{50} \leq 1/3$). In agreement with this consideration, we removed those galaxies with $\rm \Delta(R_{50})/R_{50}$ greater than 1/3, which may be detrimental for our study when we use the growth curves from IP16 in order to derive the aperture corrected $\rm H\alpha$ flux (as will be explained in Sect.~\ref{sec:app_corr}).

\item We considered galaxies with values for 1.5"$\rm/R_{50}$ higher than 0.3 (i.e. $\rm R_{50} \leq$ 5"), where 1.5" is the radius of the SDSS fibre. \cite{2013A&A...553L...7I} showed that the seeing may affect the Sérsic profile for values below this cut-off limit for the parameter 1.5"$\rm/R_{50}$, according to the analytical study carried out by \cite{2001MNRAS.321..269T}. The full width at half maximum (FWHM) of the point spread function (PSF) has a median value of $\rm \sim$3.6" in the CALIFA observations \citep{2013A&A...549A..87H,2013A&A...553L...7I}.

\item We did not consider galaxies whose SDSS spectra were classified as quasi stellar objects (QSO). In the original catalogue we have found 14,476 QSO galaxies.

\item Spectroscopic redshift in the range $\rm 0.005 \leq z \leq 0.22$. The lower z-limit of 0.005 was adopted in order to: a) minimise the effect in the photometric measurements for the larger nearby galaxies; b) to include galaxies with the lowest luminosities \citep{2004MNRAS.351.1151B}. It is expected that a sizeable fraction of the low luminosity SDSS galaxies could remain unobserved for the redshift range considered \cite[e.g.][]{2005ApJ...631..208B}.

\end{enumerate}
Our resulting primary sample contains 655,734 galaxies (74.97\% of the original catalogue).

%__________________________________________________________________
%% 2.2. Selection of the star-forming galaxies %%%%%%%%%%%%%%%%%%%%%%
\subsection{Selection of the star-forming galaxies}
\label{sec:select_sf}
\begin{figure}
  \includegraphics[width=\columnwidth]{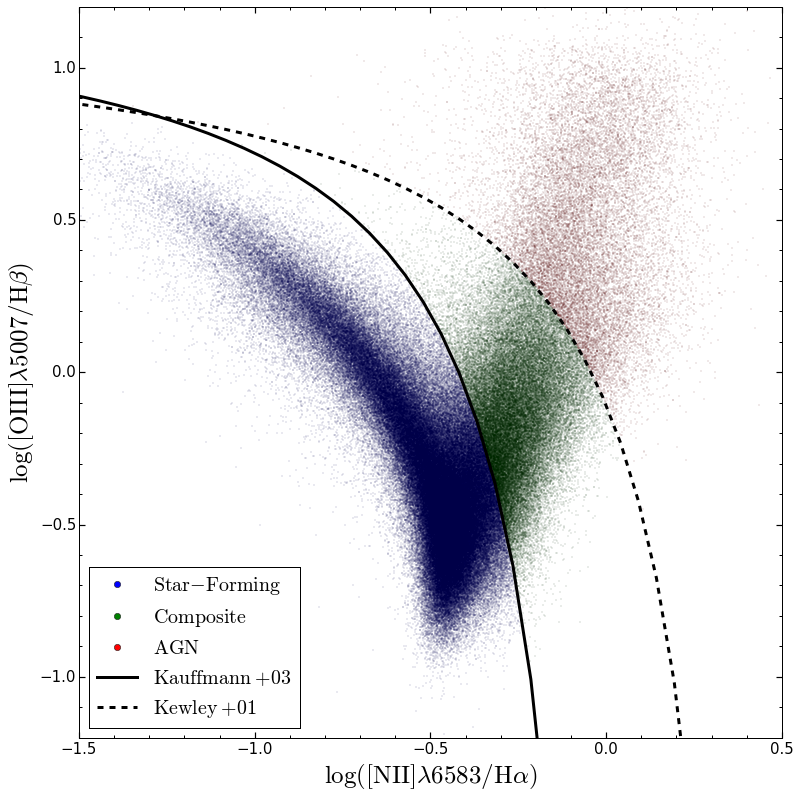}
    \caption{$\rm [OIII]\lambda\,5007/H\beta$ versus $\rm [NII]\lambda\,6583/H\alpha$ diagnostic diagram (BPT) for the SDSS galaxies. Blue, green, and red points represent star-forming galaxies in the present work (209,276), composite galaxies (57,926), and AGN galaxies (19,392), respectively. The dashed line shows the \cite{2001ApJ...556..121K} demarcation and the continuous line shows the \cite{2003MNRAS.346.1055K} curve.}
  \label{Fig:BPT_diagram}
  \end{figure}

We selected a subset of 209,276 star-forming galaxies from our primary sample described in Sect.~\ref{sec:samples} (31.92\% from our primary sample in Sect.~\ref{sec:samples}) applying the following criteria:

\begin{enumerate}[i)]
\item The S/N is greater than three for the fluxes in the strong emission lines $\rm H\alpha,\ H\beta,\ [O{III}],\ and\ [\rm N{II}]$. We have defined the S/N as the ratio of the flux and the statistical error flux, calculated with the pipeline described in \cite{2004ApJ...613..898T} \citep{2008ApJ...681.1183K}. Below this limit in S/N, an important fraction of galaxies present negative line fluxes \citep{2004MNRAS.351.1151B}.

\item We selected the pure star-forming galaxies according to the Kauffmann definition in the BPT diagnostic diagram \citep[e.g.][]{1981PASP...93....5B,1987ApJS...63..295V,2001ApJ...556..121K,2003MNRAS.346.1055K}: $\rm [OIII]\lambda\,5007/H\beta$ versus $\rm [NII]\lambda\,6583/H\alpha$ (see Fig.~\ref{Fig:BPT_diagram}): 
\begin{equation} \label{Eq:eq_2}
\rm log([OIII]/H\beta) \leq {{0.61}\over{[log([NII]/H\alpha)-0.05]}}+1.3.
\end{equation} 
$\rm log([OIII]/H\beta)$ values below this curve imply a contribution to $\rm H\alpha$ from active galactic nuclei (AGN) less than $\rm 1\%$, \citep[i.e. discarding composite or AGN galaxies;][]{2003MNRAS.346.1055K,2004MNRAS.351.1151B}.\footnote{Galaxies classified as AGN or composite using the SDSS fibre spectra and hosting star formation throughout their discs could be missed in the construction of the final sample.}

\item The $\rm H\alpha$ equivalent width $\rm (EW(H\alpha))$\footnote{The $\rm EW(H\alpha)$ is defined as the ratio between the $\rm H\alpha$ flux and the continuum flux near to $\rm H\alpha$. For the sake of simplicity in this work we assume $\rm |EW(H\alpha)|$.} is greater than or equal to three. This condition is assumed to avoid passive galaxies as defined by \cite{2011MNRAS.413.1687C}.

\end{enumerate}

%__________________________________________________________________
%__________________________________________________________________

In Table~\ref{table:median} we present the median ($\rm \pm 1\sigma$ confidence interval)\footnote{The definition of $\rm \pm 1\sigma$ used in this work is: $\rm +1 \sigma$ = percentile 84 - median; $\rm -1 \sigma$ = median - percentile 16. } of several parameters split into six redshift bins within the range $\rm 0.005\leq z\leq0.22$. In column 1 the $\rm \Delta z$ is presented, column 2 shows the $\rm \log(M_\star/M_\odot)$, columns 3 and 4 show the S/N for the $\rm H\alpha$ emission line flux and the S/N per whole spectrum range, respectively; the $\rm EW(H\alpha)$ is presented in column 5, column 6 displays the $\rm R_{50}$, and the number of galaxies per redshift bin is quoted in column 7.

%__________________________________________________________________
\begin{table*}
\caption{Values of relevant parameters corresponding to the median ($\rm \pm 1\sigma$ confidence interval) of the distribution for the star-forming galaxies in six redshift bins up to z=0.22 in the total sample.}
\label{table:median} 
\centering
\begin{tabular}{c | c c c c c c}
\hline\hline  \\[-2ex]
(1) & (2) & (3) & (4) & (5) & (6) & (7)\\[0.5ex] 
$\rm \Delta z$ & $\rm log(M_\star/M_\odot)$ & S/N F($\rm H\alpha$) & S/N (whole spec.) & $\rm EW(H\alpha)\,[\AA]$ & $\rm R_{50}$ ["] & \# galaxies \\[0.5ex] 
\hline\\[-2ex]
0.005-0.05 & $\rm 9.34^{+0.56}_{-0.46}$ & $\rm 67.16^{+34.51}_{-29.24}$ & $\rm 12.42^{+8.29}_{-5.20}$ & $\rm 24.31^{+23.47}_{-11.56}$ & $\rm 3.11^{+1.14}_{-1.14}$ & 41,883\\[0.5ex]
0.05-0.08 & $\rm 9.93^{+0.43}_{-0.40}$ & $\rm 63.57^{+30.35}_{-23.68}$ & $\rm 11.87^{+5.94}_{-4.18}$ & $\rm 23.17^{+20.13}_{-10.42}$ & $\rm 2.63^{+1.01}_{-0.86}$ & 62,616\\[0.5ex]
0.08-0.11 & $\rm 10.25^{+0.34}_{-0.36}$ & $\rm 61.08^{+29.58}_{-22.86}$ & $\rm 11.52^{+4.81}_{-3.54}$ & $\rm 22.75^{+19.62}_{-10.07}$ & $\rm 2.31^{+0.77}_{-0.67}$ & 48,178\\[0.5ex]
0.11-0.14 & $\rm 10.48^{+0.30}_{-0.36}$ & $\rm 55.29^{+28.74}_{-21.32}$ & $\rm 11.14^{+4.05}_{-3.25}$ & $\rm 23.52^{+20.75}_{-10.43}$ & $\rm 2.10^{+0.64}_{-0.58}$ & 31,084\\[0.5ex]
0.14-0.18 & $\rm 10.62^{+0.28}_{-0.44}$ & $\rm 55.04^{+29.99}_{-22.13}$ & $\rm 10.45^{+3.64}_{-3.28}$ & $\rm 25.90^{+26.52}_{-11.96}$ & $\rm 1.92^{+0.60}_{-0.59}$ & 17,465\\[0.5ex]
0.18-0.22 & $\rm 10.70^{+0.36}_{-0.50}$ & $\rm 43.47^{+32.23}_{-20.98}$ & $\rm 8.93^{+3.45}_{-3.55}$ & $\rm 29.42^{+35.78}_{-14.69}$ & $\rm 1.69^{+0.65}_{-0.59}$ & 8,050\\[0.5ex]
\hline\\[-2ex]
0.005-0.22 & $\rm 10.11^{+0.51}_{-0.66}$ & $\rm 60.76^{+31.86}_{-24.1}$ & $\rm 11.44^{+5.59}_{-3.89}$ & $\rm 23.73^{+21.74}_{-10.8}$ & $\rm 2.39^{+1.08}_{-0.76}$ & 209,276\\[0.5ex]
\hline
\end{tabular}
\tablefoot{The columns correspond to: 
(1) range of redshift considered; 
(2) median of the $\rm log(M_\star/M_\odot)$;
(3) median of the S/N ($\rm H\alpha$ flux);
(4) median of the S/N per whole spectrum range;
(5) median of the $\rm EW(H\alpha)$ [$\rm \AA$];
(6) median of the petrosian $\rm R_{50}$ ["];
(7) number of galaxies.}
\end{table*}
%__________________________________________________________________

%__________________________________________________________________
%% 3. Empirical aperture correction and derivation of SFR %%%%%%%%%%%%%%%%%%%%%%
\section{Empirical aperture correction and derivation of the SFR}
\label{sec:metho}

%__________________________________________________________________
%% 3.1 Empirical aperture correction %%%%%%%%%%%%%%%%%%%%%%
\subsection{Empirical aperture correction}
\label{sec:app_corr}
The fact that SDSS fibres (3" diameter) cover only a limited region of a galaxy at low-redshift Universe (z < 0.22), implies that only a limited amount of $\rm H\alpha$ emission can be measured. In order to derive the total $\rm H\alpha$ flux of each galaxy in our final sample, we corrected the SDSS $\rm H\alpha$ fluxes for aperture using the empirical aperture corrections in IP16.\footnote{For those galaxies in the final sample with $\rm 1.5/R_{50} \geq 2.5$ (172 objects) aperture correction of the $\rm H\alpha$ flux was applied according to Fig. 2 in IP16.} IP16 provide aperture corrections for emission lines and line ratios in a sample of spiral galaxies from the CALIFA database. Median growth curves of $\rm H\alpha$ and $\rm H\alpha/H\beta$, up to $\rm 2.5R_{50}$, were computed to simulate the effect of observing galaxies through apertures of varying radii. The median growth curve of the $\rm H\alpha$ flux (the $\rm H\alpha/H\beta$ ratio) shows a monotonous increase (decrease) with radius, with no strong dependence on galaxy properties (i.e. inclination, morphological type, and stellar mass). The IP16 sample of CALIFA star-forming galaxies spans over the whole range in galaxy mass studied in this work (see IP16 for more details).

To strengthen the relevance of the CALIFA star-forming galaxies sample used in this work for the analysis of SDSS star-forming galaxies, a series of tests have been performed. First, the similarity between the galaxy stellar mass distributions of both samples has been statistically confirmed applying a Kolmogorov-Smirnov (KS) test, from which the following results have been obtained: $\rm D_{n1,n2} = 0.08$ and p-value $\rm = 0.36$, indicating that we cannot reject the hypothesis that both samples are drawn from the same distribution. Second, the relations between SFR--$\rm M_\star$ and galaxy size--$\rm M_\star$ for star-forming galaxies from CALIFA and SDSS samples have been analysed (see Appendix~\ref{app:appendix1}); conclusive positive results have been achieved showing how the medians of the SFR of CALIFA galaxies are consistent with the results obtained in this work for SDSS galaxies over the whole range of galaxy mass (see Fig.~\ref{appfig1}). Likewise for the galaxy size--$\rm M_\star$ relation, the plot of $\rm R_{50}$ vs. $\rm M_\star$ (see Fig.~\ref{appfig2}) shows that the medians of CALIFA galaxies are clearly consistent with the SDSS galaxies distribution.

SDSS $\rm H\alpha$ fluxes were previously corrected for extinction using the Balmer decrement as measured by the $\rm H\alpha/H\beta$ ratio ($\rm {F_{H\alpha/H\beta}}^{ap,\ corr}$). To do so, the median $\rm H\alpha/H\beta$ flux ratio growth curve normalised to the $\rm H\alpha/H\beta$ flux ratio at $\rm 2.5R_{50}$, $\rm X(\alpha\beta_{50})$, was applied to each galaxy, and the $\rm H\alpha/H\beta$ ratio was computed according to its corresponding value of $\rm 1.5/R_{50}$ following IP16 (see Eq.~\ref{Eq:eq_3}):
\begin{equation} \label{Eq:eq_3}
\rm {F_{H\alpha/H\beta}}^{ap,\ corr} = \dfrac{{F_{H\alpha/H\beta}}^{0}}{X(\alpha\beta_{50})}
\end{equation}
\noindent being\\ 
\noindent $\rm X(\alpha\beta_{50})=0.0143x^5-0.1091x^4+0.2959x^3-$$\rm 0.3125x^2+0.0274x+1.1253$,\\ 
\noindent  where $\rm x = (1.5/R_{50})$ for each galaxy, $\rm {F_{H\alpha/H\beta}}^{0}$ is the SDSS $\rm H\alpha/H\beta$ ratio.

Theoretical case B recombination was assumed (the theoretical Balmer decrement, $\rm I_{H\alpha/H\beta}=2.86$; $\rm T = 10^{4}\,K$, and low-density limit $\rm n_e \sim 10^2\,cm^{-3}$; \citealt{1989agna.book.....O,1995MNRAS.272...41S}) together with the \cite{1989ApJ...345..245C} extinction curve with $\rm R_v=A_v/E(B-V)=3.1$ \citep[][]{1994ApJ...422..158O,1998ApJ...500..525S}. 

Equation \ref{Eq:eq_4} presents the aperture-corrected $\rm H\alpha$ flux, $\rm F_{H\alpha}^{ap,\ corr}$, as a function of the $\rm H\alpha$ flux in the SDSS fibre, $\rm F_{H\alpha}^{0}$, and the $\rm X(\alpha_{50})$, the median $\rm H\alpha$ flux growth curve normalised to the $\rm H\alpha$ flux at $\rm 2.5R_{50}$:
\begin{equation} \label{Eq:eq_4}
\rm F_{H\alpha}^{ap,\ corr} = \dfrac{F_{H\alpha}^{0}}{X(\alpha_{50})}            
\end{equation}
\noindent being\\
\noindent $\rm X(\alpha_{50})=0.0037x^5+0.0167x^4-$$\rm 0.2276x^3+0.5027x^2+0.1599x$\\  
\noindent where $\rm x = (1.5/R_{50})$ for each galaxy. According to IP16, for an aperture radius of $\rm 2.5R_{50}$ an average of $\rm \sim 85\%$ of the total $\rm H\alpha$ flux of (non-AGN) spiral galaxies is enclosed.

Equation \ref{Eq:eq_5} presents the total $\rm H\alpha$ flux, $\rm F_{H\alpha}^{tot}$, corrected for aperture effects and extinction:
\begin{equation} \label{Eq:eq_5}
\rm F_{H\alpha}^{tot}=\dfrac{F_{H\alpha}^{ap,\ corr}}{10^{-0.4\,A(H\alpha)}}\\[1ex]
\end{equation}

\noindent , where the extinction in $\rm H\alpha$, $\rm A(H\alpha)=1.758\,c(H\beta)$. The reddening coefficient is $\rm c(H\beta)=-\tfrac{1}{f(H\alpha)}\log{\left(\tfrac{{F_{H\alpha/H\beta}}^{ap,\ corr}}{I_{H\alpha/H\beta}}\right)}$, and $\rm f(H\alpha)$ is the reddening curve.

%__________________________________________________________________
%% 3.2 Aperture-corrected star formation rate %%%%%%%%%%%%%%%%%%%%%%
\subsection{Aperture-corrected star formation rate}
\label{sec:sfr}
The present day SFR is defined as the stellar mass formed per unit time traced by the young stars. The SFR can be derived from the $\rm H\alpha$ luminosity, and it is parametrized as $\rm SFR=L(H\alpha)/\eta_{H\alpha}$, the ratio of observed $\rm H\alpha$ luminosity\footnote{The $\rm H\alpha$ luminosity of a galaxy is $\rm L(H\alpha)=4\pi\,d^2\,F_{H\alpha}^{tot}$, d being its luminosity distance corresponding to the SDSS spectroscopic redshift.} to the conversion factor $\rm \eta_{H\alpha}$ \citep{2004MNRAS.351.1151B}. The parameter $\rm \eta_{H\alpha}$ varies with the physical properties of the galaxy, total stellar mass, and metallicity \citep[see][]{2002MNRAS.330..876C,2003A&A...410...83H,2004MNRAS.351.1151B}.

\cite{2009ApJ...703.1672K} assumed $\rm \eta_{H\alpha}$ is a constant value, $\rm \eta_{H\alpha}=10^{41.26}\,erg/s/M_\odot/yr$, which resulted in a good typical conversion factor, though $\rm \eta_{H\alpha}$ can in fact vary as much as $\sim$0.4 dex going from the least to the most massive galaxies \citep{2004MNRAS.351.1151B}. In order to parametrize the variation of the $\rm \eta_{H\alpha}$ as a function of stellar mass, we have used the median of the $\rm \eta_{H\alpha}$ likelihood distribution for the five stellar mass ranges as shown in fig. 7 in \cite{2004MNRAS.351.1151B}: $\rm log(M_\star/M_\odot) < 8$; $\rm 8 < log(M_\star/M_\odot) < 9$; $\rm 9 < log(M_\star/M_\odot) < 10$; $\rm 10 < log(M_\star/M_\odot) < 11$; $\rm log(M_\star/M_\odot) > 11$. The relation between $\rm \eta_{H\alpha}$ and galaxy stellar mass has been parametrized through a two-order polynomial fit as presented in Eq.~\ref{Eq:eq_6}:
\begin{equation} \label{Eq:eq_6}
\rm log(\eta_{H\alpha_{adj}})=-0.011\,x^2+0.124\,x+41.107,
\end{equation}

\noindent where $\rm x=log(M_\star/M_\odot).$\\

We used the values of $\rm \eta_{H\alpha}$ from Eq.~\ref{Eq:eq_6} to calculate the $\rm SFR=L(H\alpha)/\eta_{H\alpha}$ for our galaxy sample. Hereafter, we refer to the aperture-corrected $\rm log(SFR)-log(M_\star)$ relation as the SFR--$\rm M_\star$ relation for our sample of galaxies. Also, from now on we refer to the SFR and the specific SFR ($\rm sSFR = SFR/M_\star$) derived here as the empirical aperture-corrected SFR and sSFR for star-forming galaxies.

Notwithstanding the above, possible effects associated with, for example, diffuse ionised gaseous emission, geometry, or galaxy inclination should be considered \citep[e.g.][]{2006A&A...452..413R,2009ApJ...703.1672K,2012ARA&A..50..531K,2014ApJ...792L...6V}. Although a comprehensive study of these effects could be hard to handle \citep[e.g.][]{2012ARA&A..50..531K}, recent work by \citet{2013A&A...553L...7I,2016ApJ...826...71I} concluded that the $\rm H\alpha$ flux growth curve and the $\rm H\alpha/H\beta$ are insensitive to the galaxy inclination for a broad range of b/a (i.e. galaxy diameters ratio) including from edge-on to face-on galaxies. The effects of the geometry of the HII regions and the contribution of the diffuse ionised gas component are more difficult to quantify. We believe that these effects should be statistically minimised in this work, given the size of the sample and the large diversity of galaxy types used in IP16.

%__________________________________________________________________
%%%% 4. Results %%%%%%%%%%%%%%%%%%%%%%
\section{Results}
\label{sec:results}
%__________________________________________________________________
%%%% 4.1. Extinction for the SDSS star-forming galaxies sample %%%%%%%%%%%%%%%%%%%%%%
\subsection{Extinction of the star-forming galaxies in the SDSS sample}
\label{sec:resul_extinc}
\begin{figure*}
  \includegraphics[width=\textwidth]{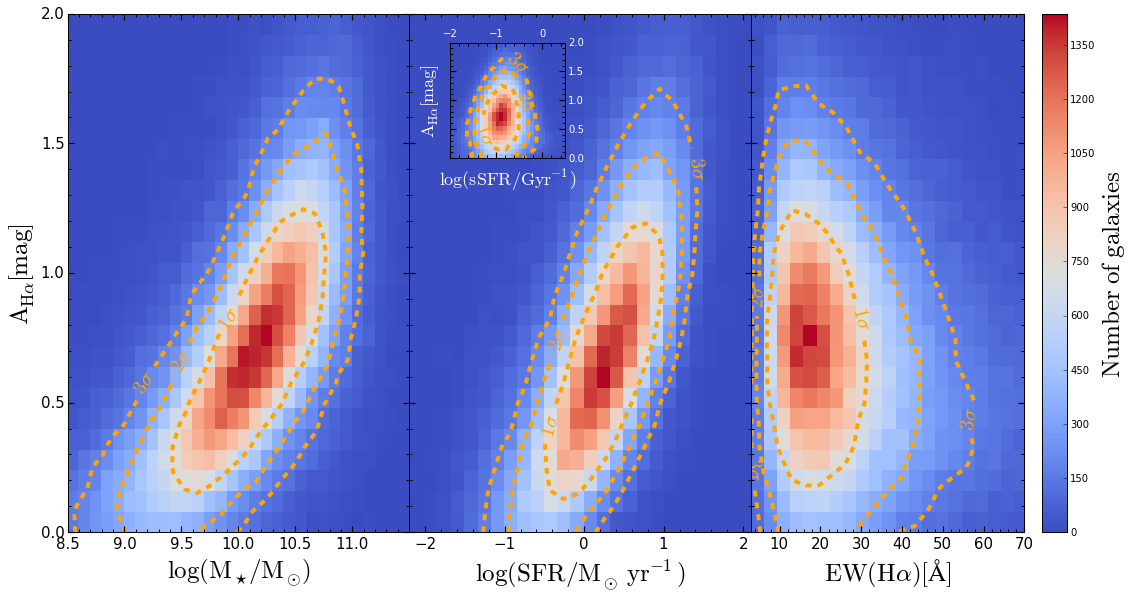}
    \caption{Density plots for the SDSS star-forming galaxies: i) Left panel: the relation between $\rm log(M_\star/M_\odot)$ and $\rm A(H\alpha)$; ii) Central panel: the relation between $\rm log(SFR/M_\odot\,yr^{-1})$ and $\rm A(H\alpha)$, the inset plot shows the $\rm A(H\alpha)$-$\rm log(sSFR/Gyr^{-1})$ relation; iii) Right panel: the relation between $\rm EW(H\alpha)$ and $\rm A(H\alpha)$. The dashed lines represent the $\rm 1\sigma$, $\rm 2\sigma$, and $\rm 3\sigma$ contours.}
  \label{Fig:EW_ahalpha}
  \end{figure*}

The extinction suffered by the $\rm H\alpha$ photons, $\rm A(H\alpha)$, derived in this work goes from 0 to 2 mag, with a median of $\rm A(H\alpha) = 0.85\,mag$\footnote{$\sim$ 4\% of galaxies in our sample present negative values of $\rm A(H\alpha)$, though consistent with $\rm A(H\alpha)=0$ mag within the errors. For these galaxies the value of $\rm A(H\alpha)$ has been set to zero.} \citep[in agreement with previous work:][]{2002A&A...383..801B,2003ApJ...599..971H,2004MNRAS.351.1151B,2004AJ....127.2511N,2013AJ....145...47M}. $\rm A(H\alpha)$ presents a strong dependence with galaxy mass, as shown in Fig.~\ref{Fig:EW_ahalpha} (left panel), being larger for more massive galaxies, \cite[see][]{2004MNRAS.351.1151B,2012ApJ...754L..29W,2013AJ....145...47M,2015MNRAS.453..879K}. Moreover, $\rm A(H\alpha)$ shows a trend with the SFR, with $\rm A(H\alpha)\leq0.2\,mag$ for those galaxies presenting the lowest SFR ($\rm \log(SFR/M_\odot\,yr^{-1}) \approx -0.5$) whereas the highest values of $\rm A(H\alpha)$ are associated with the galaxies with the largest SFR (Fig.~\ref{Fig:EW_ahalpha} central panel). Only a mild trend can be appreciated when $\rm A(H\alpha)$ is plotted versus the sSFR, confirming the strong weight that galaxy stellar mass has in the variation of $\rm A(H\alpha)$. The behaviour of $\rm A(H\alpha)$ as a function of $\rm EW(H\alpha)$ is shown in Fig.~\ref{Fig:EW_ahalpha} (right panel). Galaxies showing the median value of $\rm A(H\alpha)$ cluster around $\rm EW(H\alpha)\sim 15\AA$. For $\rm EW(H\alpha)>30\AA$, $\rm A(H\alpha)$ presents a strong decline, approximating to values near 0.2 mag and lower for $\rm EW(H\alpha)$ above $\rm 60\AA$. An upper envelope for the maximum of the $\rm A(H\alpha)$ can be seen, decreasing towards larger values of $\rm EW(H\alpha)$.

The results presented above tell us that the $\rm A(H\alpha)$ extinction correction can be substantial and therefore must be applied to all galaxy samples to be used in the study of the SFR. Moreover, these results also show that the behaviour of $\rm A(H\alpha)$ appears to be different depending on the $\rm EW(H\alpha)$ of star-forming galaxies, showing a strong relation with galaxy mass and total SFR \citep[see e.g.][]{2015MNRAS.453..879K}.

%__________________________________________________________________
%__________________________________________________________________

%__________________________________________________________________
%%%% 4.2. Parameterising the empirical SFR--M* relation for star-forming galaxies %%%%%%%%%%%%%%%%%%%%%%
\subsection{Parametrizing the aperture-corrected SFR--$\it M_\star$ relation for star-forming galaxies}
\label{sec:SFR_mass_our_result}
\begin{figure}
  \includegraphics[width=\columnwidth]{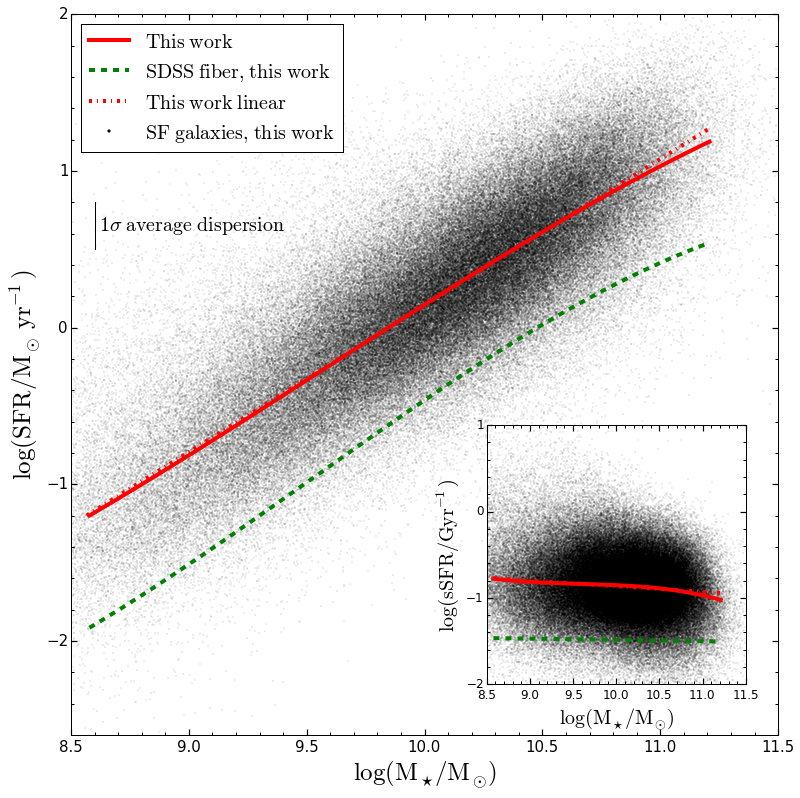}
    \caption{Relation between the SFR and $\rm M_\star$ for star-forming galaxies. The red solid line and green dashed line represent the fit to the running median for bins of 2,000 objects in this work and in the SDSS fibre, respectively. The red dotted line represents the linear fit to the running median for bins of 2,000 objects in this work. The vertical black line shows the $\rm 1\sigma$ average dispersion ($\rm \sim0.3\,dex$). The inset plot shows the sSFR--$\rm M_\star$ relation for our sample and the running median, for bins of 2,000 objects, for the sSFR corrected for aperture (red solid line) and inside the SDSS fibre (green dashed line).}
  \label{Fig:SFR_our}
  \end{figure}

In Fig.~\ref{Fig:SFR_our} we present the aperture-corrected SFR--$\rm M_\star$ relation for our sample of star-forming galaxies (see Sect.~\ref{sec:select_sf}). The SDSS fibre flux for each galaxy was corrected for aperture and extinction as explained in Sect.~\ref{sec:app_corr}. The SFR was derived for each galaxy of the sample applying the methodology described in Sect.~\ref{sec:sfr}. The running median of the distribution of points plotted in Fig.~\ref{Fig:SFR_our} (red solid line) has been fitted with the following analytical expression:
\begin{equation} \label{Eq:eq_7}
\rm log(SFR(H\alpha)) = -0.03105x^3+0.892x^2-7.571x+17.71,
\end{equation}

\noindent where $\rm x = log(M_\star/M_\odot)$. \\

For the sake of comparison with previous work (see Sect.~\ref{sec:discu}), a straight linear fit to the running median distribution in Fig.~\ref{Fig:SFR_our} has been obtained as follows:
\begin{equation} \label{Eq:eq_8}
\rm log(SFR)=\alpha x+\beta,
\end{equation}

\noindent where slope $\rm \alpha=0.935(\pm 0.001)$, $\rm \beta=-9.208(\pm 0.001)$, and $\rm x = log(M_\star/M_\odot)$.\\

Figure~\ref{Fig:SFR_our} also shows the three-order polynomial fit to the running median of the values of SFR computed from the SDSS fibre flux without any correction by aperture (green dashed line). It is clear from Fig.~\ref{Fig:SFR_our} that the corrected SFR is, on average, more than 0.65 dex above the values corresponding to the SDSS fibre flux. The SFR--$\rm M_\star$ relation obtained is consistent with the relation shown by \cite{2015A&A...584A..87C} for their sample of star-forming galaxies of the CALIFA survey (see Fig.~\ref{appfig1}).

The inset plot shows the sSFR--$\rm M_\star$ relation for our sample, corrected for aperture and extinction. The fits to the running medians of the aperture-corrected sSFR--$\rm M_\star$ (red solid line) and the SDSS fibre fluxes (green dashed line) are also shown. From the sSFR--$\rm M_\star$ relation, we observe that the fit to the running median distribution is decreasing slightly for the entire stellar mass range selected, consistent with recent predictions \citep[see e.g.][and references therein]{2015MNRAS.447.3548S}.

For the sake of completeness, in Table~\ref{table:number_sf_galaxies} we present median values ($\rm \pm 1\sigma$ confidence interval) of the derived $\rm \log(SFR)$ for twelve stellar mass bins ($\rm \Delta\log(M_\star/M_\odot)=0.25$ dex each) within the range $\rm 8.5\leq\log(M_\star/M_\odot)\leq11.5$. In column 1 the range of $\rm log(M_\star/M_\odot)$ is presented, column 2 shows the $\rm log(SFR/M_\odot\,yr^{-1})$, and the number of galaxies per stellar mass bin is quoted in column 3.

%__________________________________________________________________
%__________________________________________________________________

%__________________________________________________________________
\begin{table}
\caption{Values of aperture-corrected SFR corresponding to the median ($\rm \pm 1\sigma$ confidence interval) of the distribution for the star-forming galaxies in twelve stellar mass bins in the total sample.}
\label{table:number_sf_galaxies}
\centering
\begin{tabular}{c c c}
\hline\hline   \\[-2ex]
(1) & (2) & (3) \\[0.5ex]
$\rm \Delta log(M_\star/M_\odot)$ & $\rm \log(SFR/M_\odot\,yr^{-1})$ & \# galaxies\\[0.5ex]
\hline \\[-2ex]
8.50-8.75 & $\rm -1.11^{+0.56}_{-0.42}$ & 3,926 \\[0.5ex]
8.75-9.00 & $\rm -0.94^{+0.49}_{-0.40}$ & 6,724 \\[0.5ex]
9.00-9.25 & $\rm -0.69^{+0.43}_{-0.40}$ & 10,397 \\[0.5ex]
9.25-9.50 & $\rm -0.45^{+0.41}_{-0.38}$ & 15,405 \\[0.5ex]
9.50-9.75 & $\rm -0.19^{+0.38}_{-0.36}$ & 22,512 \\[0.5ex]
9.75-10.00 & $\rm 0.03^{+0.36}_{-0.34}$ & 29,989 \\[0.5ex]
10.00-10.25 & $\rm 0.26^{+0.35}_{-0.34}$ & 35,854 \\[0.5ex]
10.25-10.50 & $\rm 0.49^{+0.33}_{-0.35}$ & 36,230 \\[0.5ex]
10.50-10.75 & $\rm 0.71^{+0.33}_{-0.35}$ & 28,016 \\[0.5ex]
10.75-11.00 & $\rm 0.92^{+0.33}_{-0.37}$ & 14,986 \\[0.5ex]
11.00-11.25 & $\rm 1.08^{+0.33}_{-0.37}$ & 4,563 \\[0.5ex]
11.25-11.50 & $\rm 1.19^{+0.38}_{-0.40}$ & 674 \\[0.5ex]
\hline\\[-2ex]
8.50-11.50 & $\rm 0.23^{+0.58}_{-0.69}$ & 209,276 \\[0.5ex]
\hline
\end{tabular}
\tablefoot{The columns correspond to: 
(1) range of $\rm log(M_\star/M_\odot)$; 
(2) median value of $\rm \log(SFR/M_\odot\,yr^{-1})$ and error value; 
(3) number of galaxies.}
\end{table}
%__________________________________________________________________

%__________________________________________________________________
%%%% 4.3. Aperture-corrected SFR as a function of the M* and z interval %%%%%%%%%%%%%%%%%%%%%%
\subsection{Aperture-corrected SFR as a function of the $\it M_\star$ and z interval}
\label{sec:redshift_sfr_mass}
\begin{figure}
  \centering
  \includegraphics[width=\columnwidth]{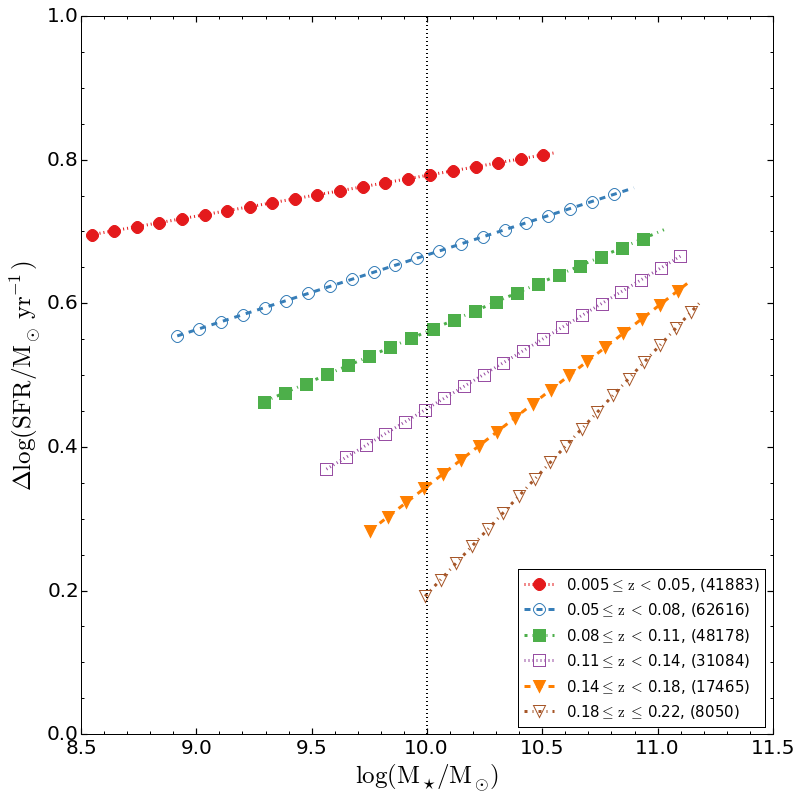}
    \caption{The empirical SFR corrections (total - SDSS fibre) vs. $\rm M_\star$ relation. Red, blue, green, magenta, orange, and gold lines represent the fit to the running median for bins of 1,000 objects in six redshift bins up to $\rm z=0.22$ for a sample of star-forming galaxies. The numbers of star-forming galaxies in each redshift bin appear on the legend. The vertical black dotted line corresponds to the reference value at $\rm log(M_\star/M_\odot)=10$.}
  \label{Fig:SFR_our_redsh}
  \end{figure}

The aperture-corrected SFR as a function of the $\rm M_\star$ is studied in six redshift bins in the range $\rm 0.005\leq z\leq 0.22$.
Figure~\ref{Fig:SFR_our_redsh} presents this evolution for the star-forming galaxy sample in the following redshift intervals: $\rm 0.005\leq z < 0.05$, $\rm 0.05\leq z < 0.08$, $\rm 0.08\leq z < 0.11$, $\rm 0.11\leq z < 0.14$, $\rm 0.14\leq z < 0.18$, and $\rm 0.18\leq z\leq 0.22$. In this figure, for each z range, the corresponding line is the fit to the running median of the differences between the aperture-corrected SFR and the SFR within the SDSS fibre as a function of the $\rm M_\star$. 

This figure shows that: i) over the whole range of galaxy stellar masses studied, the average aperture correction goes from $\sim$0.7 to $\sim$0.8 for $\rm 0.005 \leq z \leq 0.05$; ii) aperture corrections increase with galaxy mass for each redshift interval for $\rm 0.05 \leq z \leq 0.22$; reaching from $\sim$0.2 for $\rm \log(M_\star/M_\odot)=10$ to $\sim$0.6 for $\rm \log(M_\star/M_\odot)=11$, for $\rm 0.18 \leq z \leq 0.22$.

%__________________________________________________________________
%__________________________________________________________________

\begin{figure*}
  \includegraphics[width=\textwidth]{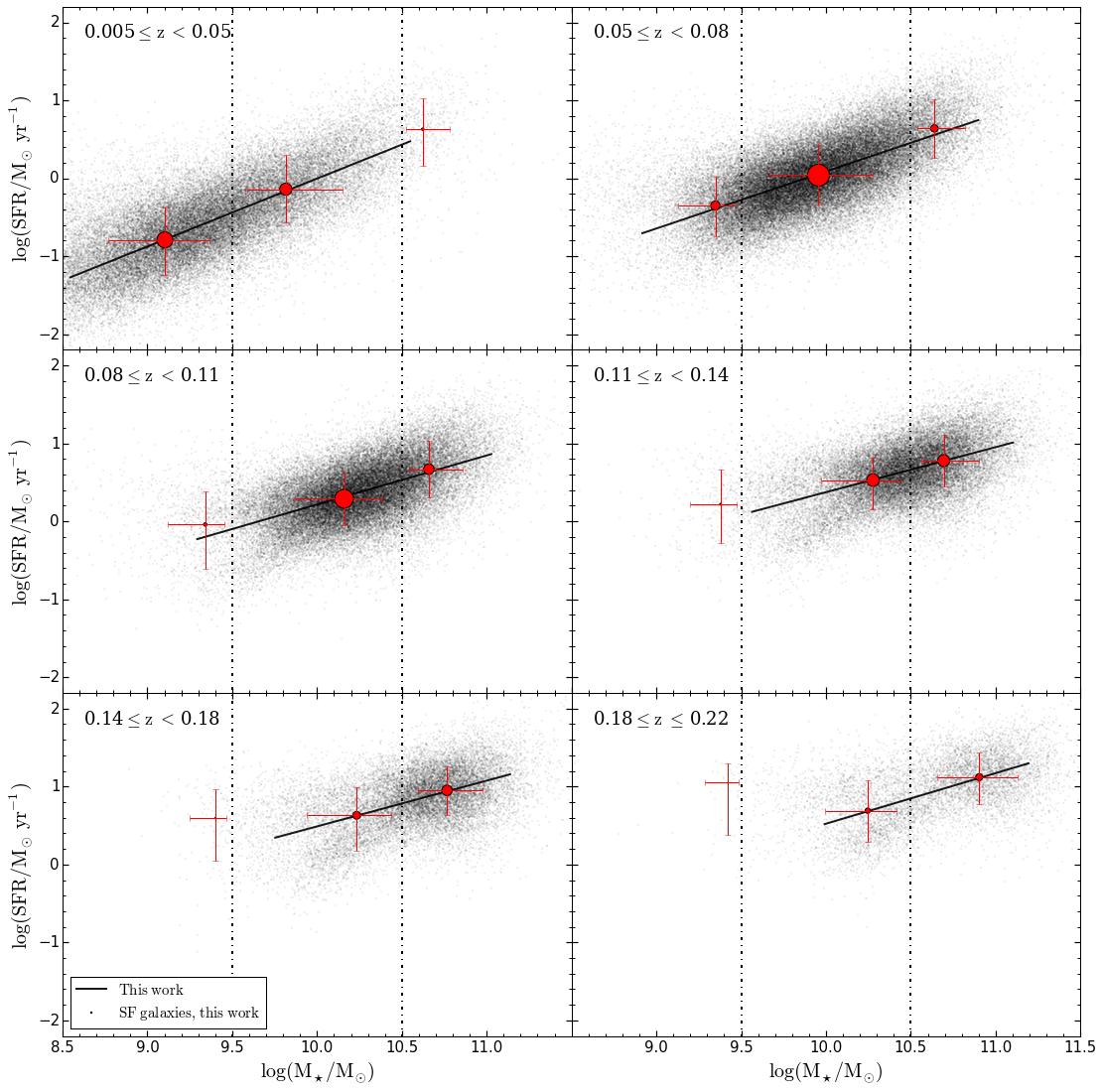}
    \caption{Relation between the SFR and $\rm M_\star$ for SDSS star-forming galaxies, along three stellar mass bins for the whole mass range ($\rm 8.5\leq\log(M_\star/M_\odot)\leq11.5$) and for six redshift ranges in the range $\rm 0.005\leq z \leq0.22$. The black solid line represents the fit to the running median for bins of 1,000 objects in this work for each redshift range. Red dots represent the medians of SFR and $\rm M_\star$ per stellar mass bin and redshift range. The error bars in x- and y-axis represent the $\rm \pm 1\sigma$ confidence interval for stellar mass and SFR, respectively. The symbol sizes increase with the number of objects contained in each mass range.}
  \label{Fig:SFR_redshift_our}
  \end{figure*}

In Table~\ref{table:number_sf_galaxies_red}, for each z range, we present the median ($\rm \pm 1\sigma$ confidence interval) of the derived $\rm \log(SFR/M_\odot\,yr^{-1})$ and $\rm \log(M_\star/M_\odot)$ along three stellar mass bins and for the whole mass range ($\rm 8.5\leq\log(M_\star/M_\odot)\leq11.5$) for six redshift bins in the range $\rm 0.005\leq z \leq0.22$. In column 1 the $\rm \Delta z$ is displayed, the $\rm \Delta log(M_\star/M_\odot)$ is presented in column 2, column 3 shows the $\rm log(SFR/M_\odot\,yr^{-1})$, column 4 shows the $\rm \log(M_\star/M_\odot)$, and the number of galaxies per stellar mass bin is quoted in column 5. All this information is presented in Fig.~\ref{Fig:SFR_redshift_our} for the sample of galaxies in the range of redshift and stellar mass considered. In this figure we show the relation between total SFR and $\rm M_\star$ for our sample of star-forming galaxies, along three stellar mass bins for the whole mass range ($\rm 8.5\leq\log(M_\star/M_\odot)\leq11.5$), and for six redshift bins in the range $\rm 0.005\leq z \leq0.22$.

%__________________________________________________________________
\begin{table}
\caption{Values of stellar mass and aperture-corrected SFR per stellar mass and redshift bins corresponding to the median ($\rm \pm 1\sigma$ confidence interval) of the distribution for the star-forming galaxies in the total sample.}
\label{table:number_sf_galaxies_red}
\centering
\begin{tabular}{c c c c r}
\hline\hline   \\[-2ex]
(1) & (2) & (3) & (4) & (5)\\[0.5ex]
$\rm \Delta z$ & $\rm \Delta \log(M_\star)$ & $\rm \log(SFR)$ & $\rm \log(M_\star)$ & \# galaxies\\[0.5ex]
\hline \\[-2ex]
 & 8.50-9.50 & $\rm -0.79^{+0.42}_{-0.45}$ & $\rm  9.10 ^{+ 0.27 }_{- 0.33 }$ & 25,872 \\[0.5ex]
0.005-0.05 & 9.50-10.50 & $\rm -0.14^{+0.44}_{-0.42}$ & $\rm  9.82 ^{+ 0.34 }_{- 0.23 }$ & 15,354 \\[0.5ex]
 & 10.50-11.50 & $\rm 0.63^{+0.39}_{-0.47}$ & $\rm  10.62 ^{+ 0.16 }_{- 0.10 }$ & 657 \\[0.5ex]
 & 8.50-11.50 & $\rm -0.55^{+0.58}_{-0.54}$ & $\rm  9.34 ^{+ 0.56 }_{- 0.46 }$ & 41,883  \\[0.5ex] \hline\\[-2ex]
 & 8.50-9.50 & $\rm -0.35^{+0.37}_{-0.39}$ & $\rm  9.35 ^{+ 0.11 }_{- 0.22 }$ & 8,775 \\[0.5ex]
0.05-0.08 & 9.50-10.50 & $\rm 0.04^{+0.39}_{-0.37}$ & $\rm  9.96 ^{+ 0.32 }_{- 0.29 }$ & 48,229 \\[0.5ex]
 & 10.50-11.50 & $\rm 0.64^{+0.37}_{-0.38}$ & $\rm  10.64 ^{+ 0.18 }_{- 0.10 }$ & 5,612 \\[0.5ex]
 & 8.50-11.50 & $\rm 0.03^{+0.45}_{-0.42}$ & $\rm  9.93 ^{+ 0.43 }_{- 0.40 }$ & 62,616  \\[0.5ex] \hline\\[-2ex]
 & 8.50-9.50 & $\rm -0.04^{+0.42}_{-0.57}$ & $\rm  9.34 ^{+ 0.11 }_{- 0.22 }$ & 1,265 \\[0.5ex]
0.08-0.11 & 9.50-10.50 & $\rm 0.29^{+0.33}_{-0.34}$ & $\rm  10.16 ^{+ 0.22 }_{- 0.29 }$ & 35,641 \\[0.5ex]
 & 10.50-11.50 & $\rm 0.67^{+0.36}_{-0.36}$ & $\rm  10.66 ^{+ 0.20 }_{- 0.12 }$ & 11,272 \\[0.5ex]
 & 8.50-11.50 & $\rm 0.36^{+0.39}_{-0.37}$ & $\rm  10.25 ^{+ 0.34 }_{- 0.36 }$ & 48,178  \\[0.5ex] \hline\\[-2ex]
 & 8.50-9.50 & $\rm 0.22^{+0.45}_{-0.50}$ & $\rm  9.38 ^{+ 0.09 }_{- 0.18 }$ & 351 \\[0.5ex]
0.11-0.14 & 9.50-10.50 & $\rm 0.53^{+0.29}_{-0.37}$ & $\rm  10.28 ^{+ 0.16 }_{- 0.31 }$ & 16,143 \\[0.5ex]
 & 10.50-11.50 & $\rm 0.78^{+0.33}_{-0.33}$ & $\rm  10.69 ^{+ 0.21 }_{- 0.14 }$ & 14,590 \\[0.5ex]
 & 8.50-11.50 & $\rm 0.64^{+0.34}_{-0.37}$ & $\rm  10.48 ^{+ 0.30 }_{- 0.36 }$ & 31,084  \\[0.5ex] \hline\\[-2ex]
 & 8.50-9.50 & $\rm 0.60^{+0.37}_{-0.55}$ & $\rm  9.40 ^{+ 0.07 }_{- 0.15 }$ & 136 \\[0.5ex]
0.14-0.18 & 9.50-10.50 & $\rm 0.63^{+0.36}_{-0.45}$ & $\rm  10.23 ^{+ 0.20 }_{- 0.30 }$ & 6,292 \\[0.5ex]
 & 10.50-11.50 & $\rm 0.95^{+0.31}_{-0.31}$ & $\rm  10.77 ^{+ 0.21 }_{- 0.17 }$ & 11,037 \\[0.5ex]
 & 8.50-11.50 & $\rm 0.85^{+0.34}_{-0.41}$ & $\rm  10.62 ^{+ 0.28 }_{- 0.44 }$ & 17,465  \\[0.5ex] \hline\\[-2ex]
 & 8.50-9.50 & $\rm 1.06^{+0.24}_{-0.68}$ & $\rm  9.42 ^{+ 0.07 }_{- 0.13 }$ & 53 \\[0.5ex]
0.18-0.22 & 9.50-10.50 & $\rm 0.69^{+0.39}_{-0.40}$ & $\rm  10.25 ^{+ 0.17 }_{- 0.25 }$ & 2,931 \\[0.5ex]
 & 10.50-11.50 & $\rm 1.12^{+0.31}_{-0.34}$ & $\rm  10.90 ^{+ 0.23 }_{- 0.25 }$ & 5,066 \\[0.5ex]
 & 8.50-11.50 & $\rm 0.99^{+0.36}_{-0.46}$ & $\rm  10.70 ^{+ 0.36 }_{- 0.50 }$ & 8,050  \\[0.5ex]
\hline\\[-2ex]
\end{tabular}
\tablefoot{The columns correspond to: 
(1) range of $\rm z$ considered; 
(2) range of $\rm \log(M_\star/M_\odot)$; 
(3) median value of the $\rm \log(SFR/M_\odot\,yr^{-1})$ and error value; 
(4) median value of the $\rm \log(M_\star/M_\odot)$ and error value;
(5) number of galaxies.}
\end{table}
%__________________________________________________________________

%__________________________________________________________________
%__________________________________________________________________

%__________________________________________________________________
%%%% 5. Discussion %%%%%%%%%%%%%%%%%%%%%%
\section{Discussion}
\label{sec:discu}
We start by comparing our SFR values (Sect.~\ref{sec:sfr}) with the ones provided by the MPA-JHU database.\footnote{A total of 4,566 star-forming galaxies from our final sample are not considered since the MPA-JHU database provided -9999 for their SFR values.} The MPA-JHU database gives the SFR within the 3" fibre of the SDSS and the total SFR corrected for aperture \citep{2004MNRAS.351.1151B,2007ApJS..173..267S} (hereafter $\rm SFR_{MPA}$). \cite{2004MNRAS.351.1151B} originally corrected fibre SFRs from aperture effects using the resolved colour information available for each galaxy. \cite{2007ApJS..173..267S} noted an overestimation of these SFRs for galaxies with low levels of star formation, attributed to the larger contribution from dusty high-metallicity starbursts inside these galaxies.\footnote{\href{http://wwwmpa.mpa-garching.mpg.de/SDSS/DR7/sfrs.html}{\texttt{http://wwwmpa.mpa-garching.mpg.de/SDSS/DR7/sfrs.html}}} Consequently, these authors improved the \cite{2004MNRAS.351.1151B} technique \citep[see][for a detailed description]{2007ApJS..173..267S} and several studies have used these values of SFR reported by these authors \citep[e.g.][]{2012ApJ...757...54Z,2015ApJ...801L..29R}.

A different aperture correction was applied by \cite{2003ApJ...599..971H}, assuming that the $\rm H\alpha$ emission-line flux can be traced across the whole galaxy by the r-band emission. \cite{2003ApJ...599..971H} derived the total $\rm H\alpha$ flux as a function of the difference between the total petrosian r-band magnitude ($\rm r_{petro}$) and the corresponding magnitude inside the SDSS fibre ($\rm r_{fibre}$) as: $\rm F_{H\alpha}^{corr}=F_{H\alpha}^{0}\times 10^{-0.4(r_{petro}-r_{fibre})}$ \citep[see also][]{2013MNRAS.432.1217P}. For the sake of comparison, the total $\rm H\alpha$ flux and the corresponding $\rm L(H\alpha)$ and SFR for all the galaxies in our final sample was recalculated according to the aperture correction recipe from \cite{2003ApJ...599..971H} [equation B3, appendix B]. Here we have refined the \cite{2003ApJ...599..971H} recipe for $\rm H\alpha$ flux aperture correction following the methodology explained in Sect.~\ref{sec:sfr}. The only difference between our refined \cite{2003ApJ...599..971H} recipe and \cite{2003ApJ...599..971H} methodology is that the former is based on Eq.~\ref{Eq:eq_6} which converts H$\rm \alpha$ into SFR, while the latter uses a constant conversion value of 41.26.

\begin{figure*}
  \includegraphics[width=\textwidth]{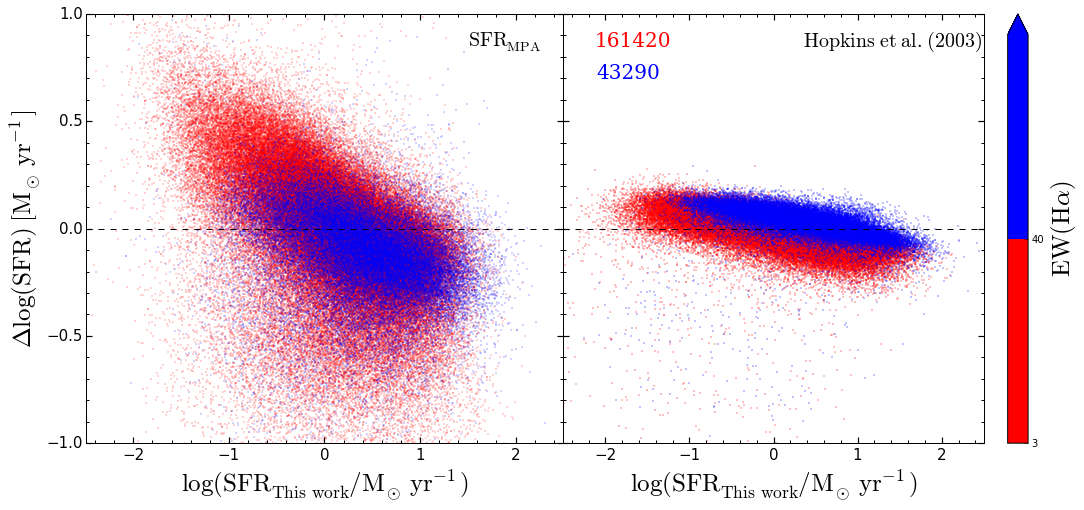}
    \caption{Left panel: difference of, $\rm \Delta\log(SFR)$, total SFR provided by MPA-JHU \citep{2004MNRAS.351.1151B,2007ApJS..173..267S} and our total empirical $\rm SFR_{This\ work}$ as a function of $\rm SFR_{This\ work}$; Right panel: difference of the total SFR derived using the \cite{2003ApJ...599..971H} aperture correction for the $\rm H\alpha$ flux and $\rm SFR_{This\ work}$ as a function $\rm SFR_{This\ work}$. Upper numbers represent the number of high- (red points) and low-level (blue points) $\rm H\alpha$ emitting galaxies. The black dashed line indicates $\rm \Delta\log(SFR)=0$.}
  \label{Fig:compare_EW}
  \end{figure*}

In Fig.~\ref{Fig:compare_EW} (left plot) we show the difference between $\rm SFR_{MPA}$ and the empirical SFR derived in this work ($\rm SFR_{This\ work}$) as a function of $\rm SFR_{This\ work}$. In the right plot we present the difference between the SFR derived applying the \cite{2003ApJ...599..971H} aperture correction and $\rm SFR_{This\ work}$ as a function of $\rm SFR_{This\ work}$. We discriminate between high- ($\rm EW(H\alpha) \geq 40\AA$; blue points) and low-level ($\rm EW(H\alpha) < 40\AA$; red points) $\rm H\alpha$ emitting galaxies. The left plot shows that: i) a significant scatter is present, especially for the low-level $\rm H\alpha$ emitters, being critical (over 1 dex) for low SFRs; and ii) those galaxies with large SFRs and high $\rm EW(H\alpha)$ show values of $\rm SFR_{MPA}$ somewhat closer to the ones derived in this work, though large systematic differences and big scatter remain. In the right plot, the average SFR values derived according to \cite{2003ApJ...599..971H} show a systematic difference with respect to $\rm SFR_{This\ work}$ ones, and a significant scatter (up to $\sim$1dex) is apparent. Galaxies with $\rm EW(H\alpha) < 40\AA$ present a larger scatter and a clear offset from those with $\rm EW(H\alpha) \geq 40\AA$. The differences between the SFR values derived using \cite{2003ApJ...599..971H} recipe and the ones obtained in this work could be reduced when we refine the \cite{2003ApJ...599..971H} methodology, as explained above; however, a noticeable scatter and some systematics between red and blue points remain. In this regard, we should bear in mind that the r-band flux of star-forming galaxies includes a contribution from $\rm H\alpha$ and nearby emission lines and, in fact, should behave as a rough tracer of HII regions. On the other hand, we note in passing that those objects with prominent bulges (likely correlating with the lower level star-formation objects in this sample) should contribute to the emission in the r-band in a more significant manner \citep[e.g.][]{2016MNRAS.455.2826R}.

\begin{figure*}
  \includegraphics[width=\textwidth]{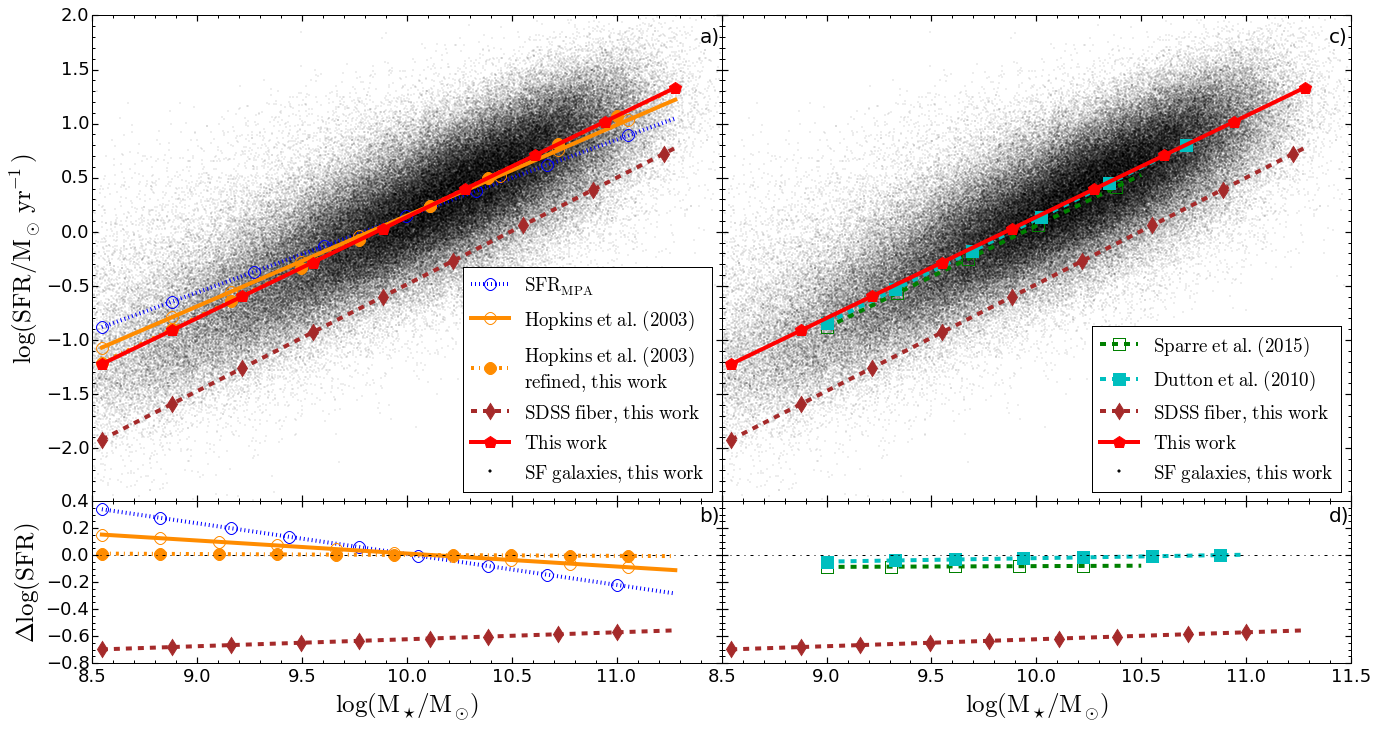}
    \caption{$\rm SFR_{This\ work}$-$\rm M_\star$ relation for star-forming galaxies compared with previous theoretical and observational works. Comparison with observational studies: Panel a) SFR--$\rm M_\star$ fits to the running median for bins of 2,000 objects obtained in this work (red solid line) and from the SDSS fibre (brown dashed line), together with the values provided by MPA-JHU (blue dotted line), using the \cite{2003ApJ...599..971H} recipe (orange solid line), and from our refined method of \cite{2003ApJ...599..971H} recipe (orange dashed line). Panel b) Difference vs. $\rm M_\star$ between SDSS fiber SFRs and $\rm SFR_{This\ work}$ and between observational studies and $\rm SFR_{This\ work}$ (colours as in upper left panel). Dotted line shows $\rm \Delta log(SFR)=0$. Comparison with theoretical studies: Panel c) SFR--$\rm M_\star$ fits to the running median for bins of 2,000 objects obtained in this work (red solid line) and from the SDSS fibre (brown dashed line), together with the predictions from \cite{2015MNRAS.447.3548S} (green dashed line) and \cite{2010MNRAS.405.1690D} (cyan dashed line). Panel d) Difference vs. $\rm M_\star$ between SDSS fiber SFRs and $\rm SFR_{This\ work}$ and between theoretical predictions and $\rm SFR_{This\ work}$ (colours as in upper right panel). Dotted line shows $\rm \Delta log(SFR)=0$.}
  \label{Fig:SFR}
  \end{figure*}

In Fig.~\ref{Fig:SFR} we show the $\rm SFR_{This\ work}$ versus $\rm M_\star$ for each galaxy in our sample and compare the fit to the SFR--$\rm M_\star$ relation derived with previous results from theoretical and observational works. We present (panel a) the linear fit to the running median to the SFR--$\rm M_\star$ relation: i) using the empirical total SFR from this work; ii) based on the $\rm SFR_{MPA}$; iii) SFR that we obtained applying the \cite{2003ApJ...599..971H} recipe and also our refined method of \cite{2003ApJ...599..971H} recipe; iv) SFR for the $\rm H\alpha$ SDSS fibre flux derived in this work. Finally, in this figure (panel c) we also show recent theoretical predictions for the SFR vs. $\rm M_\star$ relation from i) \cite{2015MNRAS.447.3548S} at z=0 using the Illustris simulation for star-forming galaxies with stellar masses between $\rm 10^9\, M_\odot$ and $\rm 10^{10.5}\, M_\odot$ and ii) by \cite{2010MNRAS.405.1690D} at z=0 using a semi-analytic model for star-forming galaxies with stellar masses between $\rm 10^9\, M_\odot$ and $\rm 10^{11}\, M_\odot$. In Fig.~\ref{Fig:SFR} (panels b and d) we present the differences between the fit to the SFR--$\rm M_\star$ relation of this work and those obtained for MPA-JHU, \cite{2003ApJ...599..971H}, our refined method of \cite{2003ApJ...599..971H} recipe, \cite{2010MNRAS.405.1690D}, and \cite{2015MNRAS.447.3548S}. The difference between the aperture-corrected SFR in this work and the SFR corresponding to the SDSS fibre is also shown.

As we can see in Fig.~\ref{Fig:SFR}, the overall difference between $\rm SFR_{MPA}$ and this work amounts to $\rm \sim0.6\,dex$ going from the less to the more massive galaxies. The fit to the SFR--$\rm M_\star$ relation shows that $\rm SFR_{MPA}$ leads to systematically larger values of the SFR for masses $\rm M_\star/M_\odot < 10^9$ ($\rm \approx$0.35 dex) and conversely, under-predicts the SFR\footnote{We have checked that the difference between $\rm SFR_{MPA}$ and our SFR values is systematic along the range $\rm 0.005 \leq z \leq 0.22$. In the larger redshift interval ($\rm 0.11 \leq z \leq 0.22$) the SFR predictions by different methods converge as expected, since the geometrical region covered by the SDSS fibre increases.} by $\rm \approx$0.3 dex for $\rm M_\star/M_\odot > 10^{11}$. Recent studies have reached similar results \cite[e.g.][]{2016MNRAS.455.2826R}.\footnote{We note that in \cite{2016MNRAS.455.2826R} the sample of galaxies with $\rm log(SFR/M_\odot\,yr^{-1}) < -1.5$ seems to be underpopulated.} In this respect, we should bear in mind that galaxies with $\rm \log(M_\star/M_\odot) \gtrsim 10.5$ may host large bulges containing little star formation, and therefore for these objects aperture correction can be problematic \citep[e.g.][]{2004MNRAS.351.1151B,2013AJ....145...47M,2016MNRAS.455.2826R}. Conversely, for less massive galaxies, the aperture problems become considerably smaller. The empirical SFR--$\rm M_\star$ fit of this work also shows a systematic difference with the relation derived using \cite{2003ApJ...599..971H}. The fit to the points obtained with the refined method of the \cite{2003ApJ...599..971H} recipe presented here gives a much better result. Excellent agreement is found with the \cite{2015MNRAS.447.3548S} and \cite{2010MNRAS.405.1690D} predictions.

It is important to note that the MS slope obtained from $\rm SFR_{MPA}$ is $\rm \sim$0.71. As far as we know, those studies that used the SFR values from the MPA-JHU catalogue obtained similar slopes \citep[e.g.][show slopes of 0.77, 0.71, 0.76, respectively]{2007A&A...468...33E,2012ApJ...757...54Z,2015ApJ...801L..29R}. On the other hand, as mentioned in Sect.~\ref{sec:SFR_mass_our_result} (see Eq.~\ref{Eq:eq_8}), the MS slope obtained in this work is 0.935. The MS slope calculated in this work is in very good agreement with the predictions of semi-analytical models by \cite{2010MNRAS.405.1690D}, giving a zero point of the MS relation consistent with our result. Agreement is also found after the comparison of the MS slope with the predictions of the Illustris cosmological hydrodynamical simulations of galaxy formation by \cite{2015MNRAS.447.3548S}. 

Finally, the SFR--$\rm M_\star$ relation obtained in this work appears consistent with recent IFS observations of modest-size samples of galaxies \citep[e.g.][]{2015A&A...584A..87C,2016MNRAS.455.2826R}. This concordance is relevant given the fact that total SFR values by \cite{2015A&A...584A..87C} and \cite{2016MNRAS.455.2826R} are obtained integrating the $\rm H\alpha$ flux over the spatially resolved galaxies, whereas in this work total SFRs are derived from the SDSS fibre $\rm H\alpha$ flux corrected for aperture effects. 

%__________________________________________________________________
%__________________________________________________________________

%__________________________________________________________________
%__________________________________________________________________

%__________________________________________________________________
%%%% 6. Summary and conclusions %%%%%%%%%%%%%%%%%%%%%%
\section{Summary and conclusions}
\label{sec:conclu}
This work provides a robust study of the total empirical SFR of galaxies and its dependence on stellar mass, extinction, and redshift. Here we present the first study that uses total $\rm H\alpha$ flux, corrected for aperture from empirical $\rm H\alpha$ growth curves, to analyse the behaviour of present-day SFR and sSFR of all SDSS star-forming galaxies. This empirical aperture correction is based on a sample of 165 spiral galaxies from the CALIFA project (IP16). Concurrently, we have derived the SFR--$\rm M_\star$ and the sSFR--$\rm M_\star$ relations applying our considerations. We have compared these relations, free from aperture effects, with other methods \citep[e.g.][]{2003ApJ...599..971H,2004MNRAS.351.1151B} and with predictions from recent theoretical models \citep[e.g.][]{2010MNRAS.405.1690D,2015MNRAS.447.3548S}. 

Our main conclusions are the following:

\begin{enumerate}[i)]
\item The mean empirical aperture-corrected SFR, averaged over galaxy stellar mass, for the entire sample of SDSS star-forming galaxies amounts to $\sim$0.65 dex.

\item The average aperture-corrected SFR for nearby galaxies, $\rm 0.005 \leq z \leq 0.05$, is between $\sim$0.7 and $\sim$0.8. For larger z ($\rm 0.05 \leq z \leq 0.22$) aperture corrections increase with galaxy mass along each redshift interval.

\item The aperture-free SFR--$\rm M_\star$ relation obtained in this work is: $\rm log(SFR)=0.935(\pm 0.001)\ log(M_\star/M_\odot)-9.208(\pm 0.001)$. When comparing total SFRs from previous works with ours, we find: a) the SFR--$\rm M_\star$ relation by MPA-JHU \citep{2004MNRAS.351.1151B,2007ApJS..173..267S} provides larger SFR values (by $\sim$0.3 dex) for $\rm \log(M_\star/M_\odot) \leq 9$; conversely, MPA-JHU SFR values for $\rm \log(M_\star/M_\odot) \geq 11$ appear systematically lower (by up to 0.3 dex); b) overall consistency is found with selected observational studies based on integral field spectroscopy of individual galaxies \citep[e.g.][]{2015A&A...584A..87C,2016MNRAS.455.2826R}; c) excellent agreement is obtained with theoretical predictions of recent semi-analytic models of disc galaxies \citep{2010MNRAS.405.1690D}, and with Illustris hydrodynamical simulations \citep{2015MNRAS.447.3548S}; d) the SFRs derived applying the \cite{2003ApJ...599..971H} recipe show a systematic difference along the range of galaxy stellar mass. When this derivation is refined following the methodology described in this work (see Sect.~\ref{sec:sfr}), together with the \cite{2003ApJ...599..971H} formula for $\rm H\alpha$ flux aperture correction, the SFRs obtained appear consistent with our results for high SFR objects, showing substantial scatter notably for the lowest SFR values. 

\item A slope $\rm d\log(SFR)/d\log(M_\star)=0.935$ is derived for our SFR--$\rm M_\star$ relation. This value is higher than those found with MPA-JHU data \citep[e.g.][]{2007A&A...468...33E,2015ApJ...801L..29R}, which are $\sim$0.76, with a significant spread. The slope found in this work perfectly agrees with recent theoretical predictions \citep{2010MNRAS.405.1690D,2015MNRAS.447.3548S}, giving further support to the empirical $\rm H\alpha$ aperture correction used. For the specific SFR (sSFR) a slightly decreasing trend is seen along the entire range of stellar mass explored.

\item The total SFR values of the entire sample present a clear correlation with extinction, in overall qualitative agreement with recent works \citep[e.g.][]{2012ApJ...754L..29W,2015MNRAS.453..879K}.

\end{enumerate}

\begin{acknowledgements}
We are grateful to Simon Verley, Maria del Carmen Argudo Fernandez, and Cristina Catalan Torrecilla for useful discussions. We thank the anonymous referee for very constructive suggestions that have improved this manuscript. SDP, JVM, JIP, CK, and EPM acknowledge financial support from the Spanish Ministerio de Economía y Competitividad under grant AYA2013-47742-C4-1-P, and from Junta de Andalucía Excellence Project PEX2011-FQM-7058. FFRO acknowledges the exchange programme Study of Emission-Line Galaxies with Integral-Field Spectroscopy (SELGIFS, FP7-PEOPLE-2013-IRSES-612701), funded by the EU through the IRSES scheme.\\

Funding for SDSS-III has been provided by the Alfred P. Sloan Foundation, the Participating Institutions, the National Science Foundation, and the U.S. Department of Energy Office of Science. SDSS-III is managed by the Astrophysical Research Consortium for the Participating Institutions of the SDSS-III Collaboration including the University of Arizona, the Brazilian Participation Group, Brookhaven National Laboratory, Carnegie Mellon University, University of Florida, the French Participation Group, the German Participation Group, Harvard University, the Instituto de Astrofísica de Canarias, the Michigan State/Notre Dame/JINA Participation Group, Johns Hopkins University, Lawrence Berkeley National Laboratory, Max Planck Institute for Astrophysics, Max Planck Institute for Extraterrestrial Physics, New Mexico State University, New York University, Ohio State University, Pennsylvania State University, University of Portsmouth, Princeton University, the Spanish Participation Group, University of Tokyo, University of Utah, Vanderbilt University, University of Virginia, University of Washington, and Yale University. The SDSS-III web site is {\tt \href{http://www.sdss3.org/}{http://www.sdss3.org}}. \\

This study makes use of the results based on the Calar Alto Legacy Integral Field Area (CALIFA) survey (\url{http://califa.caha.es/}). CALIFA is the first legacy survey being performed at Calar Alto observatory.

This research made use of Python ({\tt \href{http://www.python.org}{http://www.python.org}}) and IPython \citep{PER-GRA:2007}; Numpy \citep{2011arXiv1102.1523V}; Scipy \citep{scipyText}; Pandas \citep{mckinneyprocscipy2010}; of Matplotlib \citep{Hunter:2007}, a suite of open-source python modules that provides a framework for creating scientific plots. We also acknowledge the use of astroML \citep{astroML,astroMLText}, a Python module for machine learning and data mining built on numpy, scipy, scikit-learn, and matplotlib, and distributed under the 3-clause BSD licenseastroML. The astroML web is {\tt \href{http://www.astroml.org/}{http://www.astroml.org}}. This research made use of Astropy, a community-developed core Python package for Astronomy \citep{2013A&A...558A..33A}. The Astropy web site is {\tt \href{http://www.astropy.org/}{http://www.astropy.org}}.\\

We also acknowledge the use of STILTS and TOPCAT tools \citep{2005ASPC..347...29T}.
  
\end{acknowledgements}

\bibliography{sf_galaxies.bib}

\begin{appendix}
\section{The SFR--$\rm M_\star$ and galaxy size--$\rm M_\star$ relations of SDSS and CALIFA star-forming galaxies}
\label{app:appendix1}
\begin{figure}
\centering
\includegraphics[width=\columnwidth]{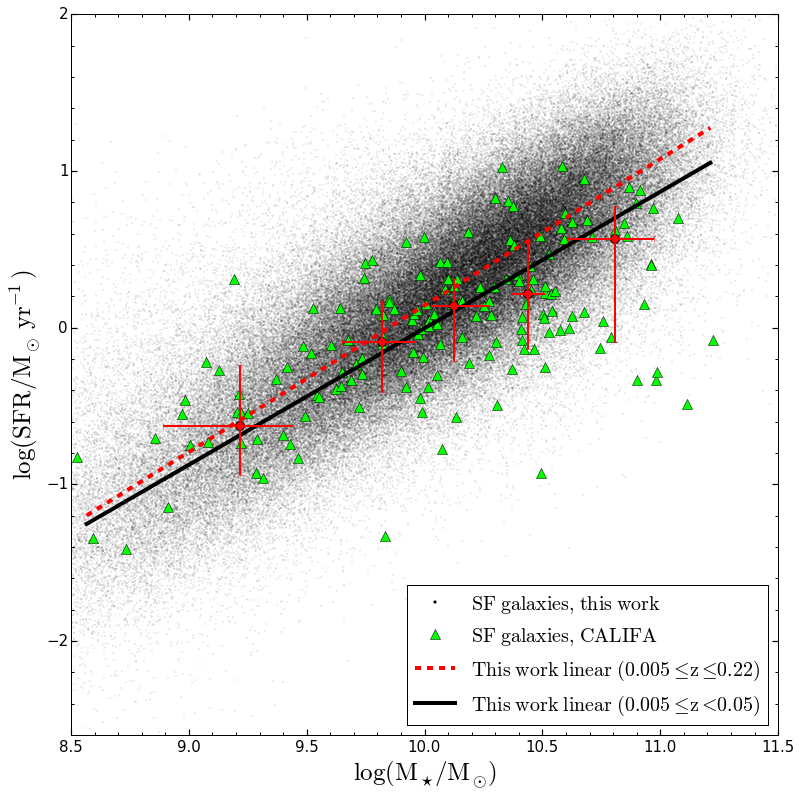}
\caption{SFR--$\rm M_\star$ relation for SDSS (this work) and CALIFA \citep{2015A&A...584A..87C} star-forming galaxies. SFR--$\rm M_\star$ fit to the running median for bins of 2,000 objects obtained in this work for the complete sample (red dashed line) and for bins of 1,000 objects in the range of z between $\rm 0.005 \leq z < 0.05$ (black solid line). Red dots represent the median values of SFR and $\rm M_\star$ for CALIFA star-forming galaxies along five stellar mass bins for an equitable number of galaxies in each bin. The error bars in x- and y-axis represent the $\rm \pm 1\sigma$ confidence interval for stellar mass and SFR, respectively.}
\label{appfig1}
\end{figure}

\begin{figure}
\centering
\includegraphics[width=\columnwidth]{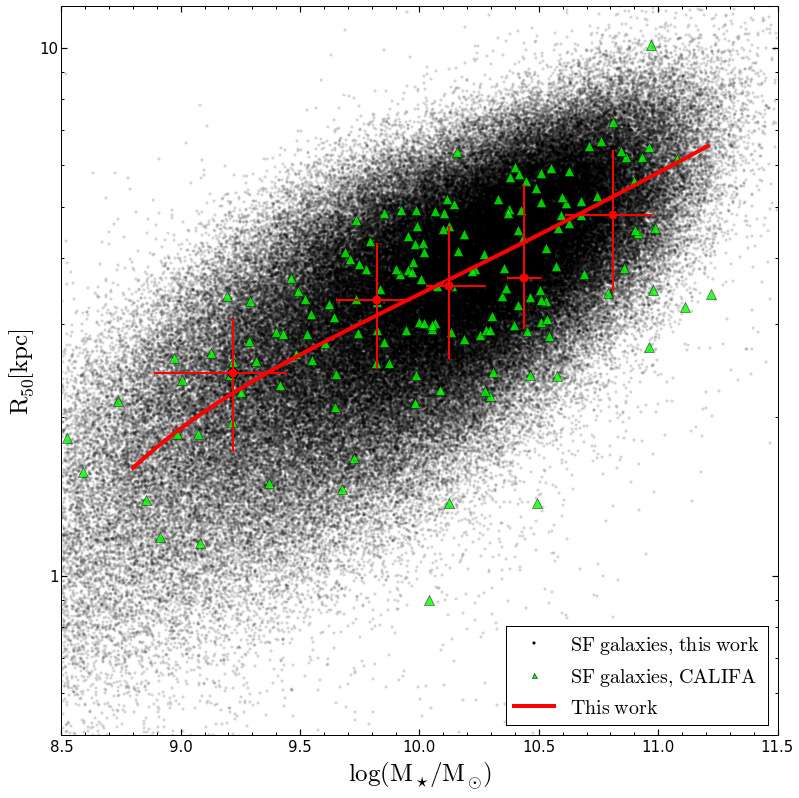}
\caption{Size--$\rm M_\star$ relation for SDSS (this work) and CALIFA \citep{2015A&A...584A..87C} star-forming galaxies. Red solid line represents the linear fit to the running median for bins of 2,000 objects in the SDSS star-forming galaxies. Red dots represent the median values of size and $\rm M_\star$ for CALIFA star-forming galaxies along five stellar mass bins for an equitable number of galaxies in each bin. The error bars in x- and y-axis represent the $\rm \pm 1\sigma$ confidence interval for stellar mass and size, respectively.}
\label{appfig2}
\end{figure}

In order to strengthen the relevance between CALIFA and SDSS galaxies, we have compared CALIFA star-forming galaxies in the SFR--$\rm M_\star$ and size--$\rm M_\star$ diagrams with the SDSS galaxies (Figs.~\ref{appfig1} and \ref{appfig2}, respectively).
Figure~\ref{appfig1} shows the SFR--$\rm M_\star$ for SDSS and CALIFA star-forming galaxies. The SFRs values for the CALIFA galaxies were computed consistently following the methodology presented in Sect.~\ref{sec:sfr} of this work, using the data for the star-forming galaxies in \citet[private communication]{2015A&A...584A..87C}. The linear fit to the running median of the aperture-corrected SFR--$\rm M_\star$ distribution for the complete sample ($\rm 0.005 \leq z \leq 0.22$), and for the first redshift range considered ($\rm 0.005 \leq z < 0.05$) are shown, together with the median values of SFR and $\rm M_\star$ along five stellar mass bins for CALIFA star-forming galaxies; equitable number of galaxies have been considered in each bin.

Figure~\ref{appfig2} shows the size--$\rm M_\star$ for SDSS and CALIFA star-forming galaxies. The linear fit to the running median of the size--$\rm M_\star$ distribution for the complete sample ($\rm 0.005 \leq z \leq 0.22$) and the median values of size and $\rm M_\star$ along five stellar mass bins for CALIFA star-forming galaxies are shown; equitable number of galaxies have been considered in each bin.

From the figures we can see that the median values of the SFR of CALIFA galaxies are representatives with the results obtained in this work for (nearby $\rm 0.005 \leq z < 0.05$) SDSS galaxies over the whole range of galaxy mass. Likewise, the median values of the galaxy size of CALIFA galaxies are visibly consistent with the SDSS galaxies distribution. Taking into account these results, we can consider CALIFA star-forming galaxies as representative of the SDSS star-forming galaxies sample used in this work.

\end{appendix}

\end{document}